\documentclass[fleqn,usenatbib]{mnras}

\usepackage[T1]{fontenc}

\DeclareRobustCommand{\VAN}[3]{#2}
\let\VANthebibliography\thebibliography
\def\thebibliography{\DeclareRobustCommand{\VAN}[3]{##3}\VANthebibliography}

\usepackage{graphicx}	% Including figure files
\usepackage{amsmath}	% Advanced maths commands
\usepackage{amssymb}	% Extra maths symbols

\newcommand{\MgII}{Mg\,{\sc ii}}
\newcommand{\CIII}{C\,{\sc iii]}}
\newcommand{\CIV}{C\,{\sc iv}}

\title{The Mass Distribution of Quasars in Optical Time-domain Surveys}

\author[Mouyuan Sun]{
Mouyuan Sun,$^{1}$\thanks{E-mail: msun88@xmu.edu.cn}
\\
$^{1}$Department of Astronomy, Xiamen University, Xiamen, Fujian 361005, China
}

\date{Accepted XXX. Received YYY; in original form ZZZ}

\pubyear{2023}

\begin{document}
\label{firstpage}
\pagerange{\pageref{firstpage}--\pageref{lastpage}}
\maketitle

\begin{abstract}

The determination of supermassive black hole (SMBH) masses is the key to 
understanding the host galaxy build-up and the SMBH mass assembly histories. 
The SMBH masses of non-local quasars are frequently estimated via the 
single-epoch virial black-hole mass estimators, which may suffer from 
significant biases. Here we demonstrate a new approach to infer the 
mass distribution of SMBHs in quasars by modelling quasar UV/optical 
variability. Our inferred black hole masses are systematically 
smaller than the virial ones by $0.3\sim 0.6$ dex; the 
$\sim 0.3$ dex offsets are 
roughly consistent with the expected biases of the virial black-hole 
mass estimators. In the upcoming time-domain astronomy era, our 
methodology can be used to constrain the cosmic evolution of quasar 
mass distributions. 

\end{abstract}



\section{Introduction}
\label{sec:intro}
The mass distribution of supermassive black holes (SMBHs) in galaxy 
centres is vital for our understanding of their mass assembly histories, 
the active galactic nucleus (AGN) feedback energy budget, and the 
evolution of host galaxies. For distant AGNs (including quasars, i.e., 
luminous AGNs with broad emission lines), their SMBH masses ($M_{\rm{BH}}$) 
are often estimated via the reverberation mapping method \citep{Blandford1982} 
and its simplified version, the single-epoch virial black-hole mass 
estimators \citep[for a review, see, e.g.,][]{Shen2013}. The two methods 
are built upon the assumption that the broad emission-line regions (BLRs), 
which are photoionized by the central engine and emit broad emission 
lines, are virialized. The virial mass estimators are calibrated 
\citep[e.g.,][]{Onken2004} with respect to the famous 
$M_{\rm{BH}}$--$\sigma$ relation \citep{Ferrarese2000, Gebhardt2000} 
under the assumption that AGNs have similar virial factors. 
It has been shown that the virial black hole mass ($M_{\rm{BH, vir}}$) 
has substantial intrinsic uncertainties ($\gtrsim 0.4$ dex) and suffers 
from systematic biases \citep[$\gtrsim 0.3$ dex; see, e.g.,][]{Shen2010}. 
Ongoing \citep[e.g.,][]{Du2016, Homayouni2020, Yu2022} and future 
reverberation mapping campaigns may considerably improve the accuracy of 
$M_{\rm{BH, vir}}$ by monitoring a large quasar sample and 
modelling their BLR structures \citep[e.g.,][]{Pancoast2014, LiSS2021}. 

In the upcoming time-domain era of astronomy, scaling relations between 
$M_{\rm{BH}}$ and AGN variability properties are often proposed to 
estimate $M_{\rm{BH}}$ since AGN variability is unambiguous in 
various wavelengths. Previous studies often focus on the X-ray 
excess variance and find that the corresponding mass estimation precision 
is comparable to the reverberation mapping method \citep[e.g.,][]{Zhou2010, 
Kelly-Treu2013}. 
Some works take an alternative approach by adopting the 
$M_{\rm{BH}}$-galaxy total stellar mass ($M_{\star}$) scaling 
relations to model the ensemble X-ray variability and constrain the mass 
distribution of AGNs in deep X-ray surveys \cite[e.g.,][]{Georgakakis2021}. 
In the next decade, the Legacy Survey of Space and Time (LSST) of the 
Vera C. Rubin Observatory will comprehensively measure AGN variability 
in the optical bands \citep[e.g.,][]{Brandt2018}. While several studies 
have discussed the possibility to use optical variability to estimate 
$M_{\rm{BH}}$ \citep[e.g.,][]{Kelly2013,Sun2020a,Burke2021}, the 
validity of such a method has not yet been well demonstrated. 

In this study, we aim to constrain the mass distribution of quasars by 
modelling their optical structure functions measured from the Sloan Digital 
Sky Survey (SDSS) Stripe 82 \citep[S82;][]{MacLeod2012} and the Panoramic 
Survey Telescope and Rapid Response System Survey 
\citep[Pan-STARRS;][]{Chambers2016, Flewelling2020} light curves. The 
manuscript is 
formatted as follows. In Section~\ref{sect:obs}, we present the observed 
quasar variability. In Section~\ref{sect:mock}, we introduce our quasar 
variability forward modelling procedures. In Section~\ref{sect:result}, 
we show the modelling results. Summary and future prospects are made in 
Section~\ref{sect:sum}.

\section{Observations}
\label{sect:obs}

Following \cite{Suberlak2021}, we use the SDSS S82 quasar light curves and 
Pan-STARRS observations to explore quasar variations on observed-frame 
timescales up to $15$ years. The SDSS S82 quasar light curves (each light 
curve typically contains $40$ data points) are downloaded from 
\url{https://faculty.washington.edu/ivezic/cmacleod/qso_dr7/Southern.html}. 
Then, we cross-match each SDSS S82 quasar with the Pan-STARRS second 
data release (\url{https://catalogs.mast.stsci.edu/panstarrs/}; each 
light curve typically has twelve observations) to extend the light-curve 
duration. 

We focus on the $2114$ quasars at redshift $1.4<z<1.8$ because the redshift 
distribution of our quasar sample peaks at this range. The median 
redshift for the selected quasars is $1.6$. We cross-match 
the selected quasars with the quasar property catalogue of 
\cite{Shen2011} to obtain their bolometric luminosities ($L_{\rm{bol}}$) 
and $M_{\rm{BH, vir}}$; our conclusions remain unchanged if 
we use the updated quasar property catalogue of \cite{Wu2022}. The selected 
quasars are divided into 
five $\log L_{\rm{bol}}$ bins; each bin has the same number of quasars. 
Hence, the $\log L_{\rm{bol}}$ bins are defined as follows: 
[$45.79, 45.97$], [$45.97, 46.10$], [$46.10, 46.25$], [$46.25, 46.47$], 
and [$46.47, 46.77$]. 

\begin{figure}
    \centering
    \includegraphics[width=\columnwidth]{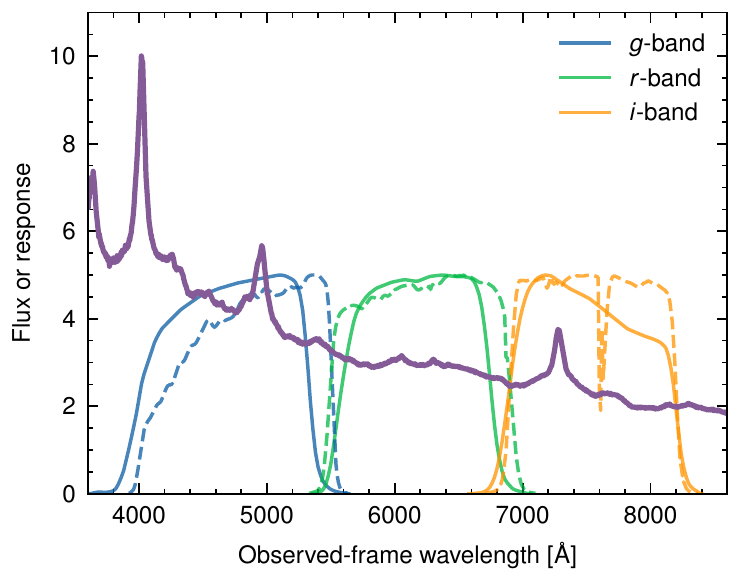}
    \caption{Transmission curves of SDSS (solid curves) 
    and Pan-STARRS (dashed curves) $gri$ bands. The purple curve 
    represents the SDSS composite spectrum at redshift 
    $z=1.6$. Contamination due to strong UV emission lines, e.g.,  
    \CIV\ and Ly$\protect \alpha$, are negligible in our studies. }
    \label{fig:bands}
\end{figure}

The SDSS S82 data have five bands, i.e., $ugriz$. The Pan-STARRS survey 
has $grizy$ filters. The SDSS $gri$ and Pan-STARRS $gri$ have similar 
filter response curves; the SDSS $z$ differ significantly from Pan-STARRS 
$z$. Hence, we focus on the SDSS and Pan-STARRS $gri$ light curves. For 
the redshift ranges of our selected quasars, according to the SDSS composite 
spectrum \citep{Berk2001}, the $r$-band flux is 
emission-line free; $g$ and $i$ fluxes have weak inevitable contamination 
due to \MgII\ and \CIII\ (Figure~\ref{fig:bands}). We assume that such 
contamination is negligible \citep[see, e.g.,][]{MacLeod2012} in our 
subsequent analysis. 

Following \cite{Suberlak2021}, we use the SDSS S82 standard stars v4.2 of 
\cite{Thanjavur2021} to find the difference ($\delta m_{\rm{diff}}$) 
of the SDSS and Pan-STARRS magnitudes at each band as a 
function of the SDSS $g-r$ colour index, 
\begin{equation}
    \label{eq:corr}
    \delta m_{\rm{diff}} = A + B(m_g-m_r) \\,
\end{equation}
where $m_g$ and $m_r$ are the apparent magnitudes in SDSS 
$g$ and $r$ bands, respectively. We use the $g-r$ colour 
index rather than the $g-i$ one because the $r$ band is emission-line 
free (Figure~\ref{fig:bands}). 
For $g$, $r$, and $i$ bands, we find that $A=0.01836$, 
$0.0009$, and $0.0199$, and $B=0.154$, $0.001$, and $0.008$, respectively. 
Then, we use the empirical correction of Eq.~\ref{eq:corr} and quasar 
$g-r$ colour indexes to convert the observed Pan-STARRS magnitudes into 
the corresponding SDSS magnitudes and construct merged quasar light curves. 

We adopt the SDSS S82 and Pan-STARRS merged light curves to calculate 
the squared structure functions ($\rm{SF}^2$) which measure 
the statistical variance of quasar magnitude fluctuations ($\Delta m$) 
at a given band and a fixed rest-frame time separation ($\Delta t$). 
We consider twenty-two $\Delta t$ which are evenly spaced between 
rest-frame $1$ day and $2, 000$ days in logarithmic space. For each 
$\Delta t$, the corresponding variability amplitude of $\Delta m$ 
is calculated via the squared normalized average absolute 
deviation (squared NAAD), 
\begin{equation}
    \label{eq:naad}
    \rm{SF}^2(\Delta t)\equiv \sigma_{\rm{NAAD}}^2 = \frac{\pi}{2} 
    \times \left(\overline{|\Delta m - \overline{\Delta m}|}\right)^2 \\,
\end{equation}
where $\overline{\Delta m}$ represents the average value of $\Delta m$; 
the factor, $\pi/2$, ensures that $\sigma_{\rm{NAAD}}^2$ is 
identical to the variance if the distribution of $\Delta m$ is 
perfectly Gaussian. We do not subtract $\Delta m$ fluctuations due to 
magnitude measurement uncertainties. Hence, on very short $\Delta t$ 
(e.g., $\sim 1$ day), $\sigma_{\rm{NAAD}}^2$ traces the measurement 
uncertainties of $\Delta m$; on timescales of $\geq 10$ days (in 
rest-frame), $\sigma_{\rm{NAAD}}^2$ probes quasar intrinsic 
variability. 

\begin{figure*}
    \includegraphics[width=2\columnwidth]{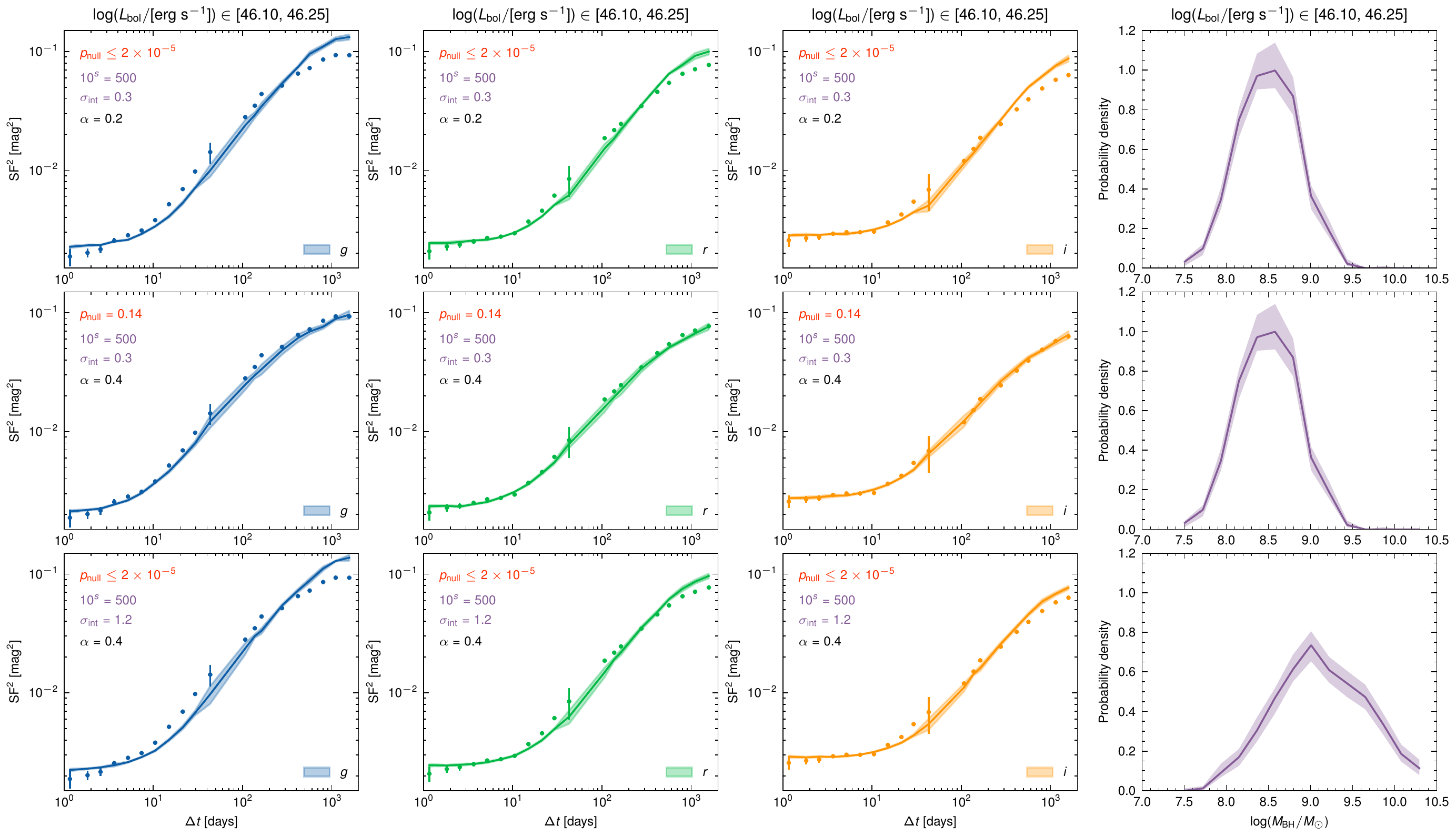}
    \caption{An illustration of the observed (points) 
    squared structure functions versus the model (solid curves) and 
    $1\sigma$ uncertainties (i.e., shaded regions) for models with 
    two different $\alpha$ values or $M_{\mathrm{BH}}$ distributions 
    (shown in the fourth column). The $M_{\mathrm{BH}}$ distributions 
    are controlled by $10^s$ and $\sigma_{\mathrm{int}}$. 
    The upper, middle, and bottom rows correspond to models with [$10^s=500$, 
    $\sigma_{\rm{int}}=0.3$ dex, $\alpha=0.2$], [$10^s=500$, 
    $\sigma_{\rm{int}}=0.3$ dex, $\alpha=0.4$], and [$10^s=500$, 
    $\sigma_{\rm{int}}=1.2$ dex, $\alpha=0.4$], respectively. The 
    first, second, and third columns represent the results for $g$, 
    $r$, and $i$ bands, respectively. Panels in the same column have 
    exactly the same observed squared structure functions whose error 
    bars are estimated via $1/N_{\mathrm{p}}$, where $N_{\mathrm{p}}$ 
    is the number of data pairs in each $\Delta t$ bin; error bars are 
    smaller than point sizes in most $\Delta t$ bins. The 
    $p_{\mathrm{null}}$ value is calculated by simultaneously 
    considering the observed squared structure functions of all 
    five luminosity bins and three bands; hence, $p_{\mathrm{null}}$ 
    is the same in each row. The model with [$10^s=500$, 
    $\sigma_{\rm{int}}=0.3$ dex, $\alpha=0.4$] fits the observations 
    better than other two models with either different $\alpha$ or 
    $\sigma_{\rm{int}}$. 
    Note that the results in this figure are for the luminosity bin 
    $46.10\leq \log (L_{\rm{bol}}/[\rm{erg\ s^{-1}}])<46.25$; the 
    results for other four luminosity bins are presented 
    in Appendix A. }
    \label{fig:sfplot}
\end{figure*}

The observed squared structure functions are shown in Figure~\ref{fig:sfplot}. 
As shown in previous SDSS S82 studies 
\citep[e.g.,][]{MacLeod2012,Suberlak2021}, the squared structure 
functions increase with $\Delta t$ and anti-correlate with wavelengths 
and quasar luminosities, which agree well with the Corona Heated Accretion 
disc Reprocessing (CHAR) model predictions \citep{Sun2020a}. It has been 
shown that the CHAR model can reproduce the ensemble structure functions of 
SDSS S82 light curves \citep{Sun2020b}.

\section{Modelling the observed structure functions}
\label{sect:mock}

\begin{table}
    \centering
    \caption{The normalization ($10^s$) and intrinsic scatter 
    ($\sigma_{\rm{int}}$) of the $M_{\rm{BH}}$-$M_{\star}$ 
    relation}
    \label{tbl:scaling}
    \begin{tabular}{ccc}
    \hline
    Case & $10^{s}$ & $\sigma_{\rm{int}}$\\
    \hline
    1 & 500 & 0.3 \\
    2 & 500 & 0.6 \\
    3 & 500 & 1.2 \\
    4 & 1000 & 0.3 \\
    5 & 1000 & 0.6 \\
    6 & 1000 & 1.2 \\
    7 & 2000 & 0.3 \\
    8 & 2000 & 0.6 \\
    9 & 2000 & 1.2 \\
    10 & 4000 & 0.3 \\
    11 & 4000 & 0.6 \\
    12 & 4000 & 1.2 \\
    13 & 8000 & 0.3 \\
    14 & 8000 & 0.6 \\
    15 & 8000 & 1.2 \\
    \hline
    \end{tabular}
\end{table}

We use the CHAR model \citep[for the implemented physics, please refer 
to][]{Sun2020a} to generate mock light curves; then, we compare the 
squared structure functions of the mock data with the observed ones. 
SMBHs are assumed to be Schwarzschild black holes, and the inner and 
outer boundaries of the accretion discs are $3\ R_{\rm{S}}$ and 
$3000\ R_{\rm{S}}$ (where $R_{\rm{S}}$ is the Schwarzschild 
radius), respectively. To use the CHAR model, we have to assign 
four parameters, namely, $L_{\rm{bol}}$, 
$M_{\rm{BH}}$, the dimensionless viscosity parameter $\alpha$, and 
the variability amplitude ($\delta_{\rm{mc}}$) of the magnetic 
fluctuations in the CHAR model. We consider two cases for $\alpha$, i.e., 
$\alpha=0.2$ and $\alpha=0.4$ \citep[e.g.,][]{King2007}. We also assume 
that $\delta_{\rm{mc}}$ is independent of $M_{\rm{BH}}$ or 
$L_{\rm{bol}}$, which are the remaining two parameters. 

Different from our previous study \citep[][which only uses SDSS S82 
light curves and probes rest-frame timescales of $\Delta t\lesssim 500$ 
days]{Sun2020b}, we now do not fix the model 
$M_{\rm{BH}}$ to $M_{\rm{BH, vir}}$ since the 
latter suffers from various biases and has substantial uncertainties 
\citep[for a review, see, e.g.,][]{Shen2013}. Instead, we take a 
forward-modelling approach and consider the 
correlation between $M_{\rm{BH}}$ and the total stellar mass 
($M_{\star}$) of their host galaxies \citep[e.g.,][and references 
therein]{Sun2015, Li2021}, i.e., 
\begin{equation}
    \label{eq:scaling}
    \log(M_{\rm{BH}}/M_{\odot}) = \log(M_{\star}/M_{\odot}) - s \\,
\end{equation}
with the intrinsic scatter follows a log-normal distribution, 
whose standard deviation is $\sigma_{\rm{int}}$. We assume 
that the probability density function of $M_{\star}$ is a summation of two 
Schechter functions, and we adopt the continuity model parameters (at redshift 
$z=1.6$) obtained by \citet[][see their appendix B]{Leja2020}. For the value 
of $s$ (i.e., the normalization of the scaling relation), we consider five 
cases, i.e., $10^s=500$, $1000$, $2000$, $4000$, and $8000$, respectively. 
For $\sigma_{\rm{int}}$, we explore three values, i.e., 
$\sigma_{\rm{int}}=0.3$ dex, $0.6$ dex, and $1.2$ dex, respectively. 
That is, we consider $5\times 3=15$ cases (for a summary, see 
Table~\ref{tbl:scaling}). It should be noted that this approach is 
equivalent to assuming a specific intrinsic distribution of 
$\log M_{\rm{BH}}$, which is a convolution of the summation of two 
Schechter functions and a normal distribution (whose mean and standard 
deviation are $s$ and $\sigma_{\rm{int}}$, respectively). Due to 
various selection effects, the $\log M_{\rm{BH}}$ distribution 
of the observed quasars are different from the intrinsic one. Below, 
we will simulate the detectable mock quasars by applying selection 
cuts to the intrinsic $\log M_{\rm{BH}}$ distribution. 

Our mock light-curve simulation process is similar to that of 
\cite{Georgakakis2021}. First, we randomly 
sample mock galaxies (with $M_{\star}$) according to the galaxy stellar 
mass function of \cite{Leja2020}. Second, for each mock galaxy, we assign 
an SMBH with $M_{\rm{BH}}$ according to Eq.~\ref{eq:scaling}. Third, 
we assume the Eddington ratio distribution of AGNs follows the Schechter 
function, i.e., 
\begin{equation}
    \frac{dt}{d\log \lambda_{\rm{Edd}}} \propto
    \left(\frac{\lambda_{\rm{Edd}}}{\lambda_{\rm{Edd, 
    cut}}}\right)^{-\gamma} 
    \exp(-\lambda_{\rm{Edd}}/\lambda_{\rm{Edd, cut}}) \\,
\end{equation}
where $\lambda_{\rm{Edd}}$ represents the Eddington ratio; the slope 
$\gamma$ and $\lambda_{\rm{Edd, cut}}$ are fixed\footnote{We also tried 
to set $\gamma=0.6$ \citep[e.g.,][]{Aird2012} and found that our conclusions 
remain largely unchanged.} to be $0.2$ and $0.4$ \citep{Jones2016}, 
respectively. The lower limit of $\log \lambda_{\rm{Edd}}$ is fixed to be 
$-4$ (i.e., much smaller than the observed 
$\log \lambda_{\rm{Edd}}$ for quasars) for Case 1; the lower limits of 
$\lambda_{\rm{Edd}}$ for other cases are adjusted to ensure that their 
corresponding mean values of $\log L_{\rm{bol}}$ are identical to that of 
Case 1. Note that our models can reproduce the observed 
quasar luminosity function. Fourth, we calculate 
the corresponding bolometric luminosity ($L_{\rm{bol}}$) and convert it 
into the $3000\ \rm{\AA}$ continuum luminosity ($L_{3000}$, i.e., $\lambda 
L_{\lambda}$ at $\lambda=3000\ \rm{\AA}$) by adopting a constant bolometric 
correction\footnote{We confirm that our results remain the same if we use the 
luminosity-dependent bolometric correction factor.} of $5$ 
(whose intrinsic scatter follows a log-normal distribution 
with the standard deviation of $0.3$), 
and obtain the corresponding $i$-band apparent magnitude ($m_i$) and the FWHM 
of \MgII\ \citep[see Eq. 8 in][]{Shen2011}. Fifth, we select mock quasars with 
$m_i$ brighter than $20.3$ mag and the FWHM of \MgII\ larger than 
$10^3\ \rm{km\ s^{-1}}$ (hereafter the ``detectable'' mock quasars); the 
$m_i$ limit of $20.3$ mag corresponds to the 95-th percentile of the $m_i$ 
distribution of the $2114$ quasars in Section~\ref{sect:obs}. Sixth, for each 
luminosity bin introduced in Section~\ref{sect:obs}, we randomly select the 
same number of mock quasars whose $L_{\mathrm{bol}}$ (i.e., $5L_{3000}$) fall 
into the luminosity bin. Seventh, for each selected mock quasar, we use its 
$L_{\rm{bol}}$ and $M_{\rm{BH}}$ and the 
CHAR model to calculate their multi-wavelength light curves. The mock light 
curves share the same sampling pattern as real observations. In this procedure, 
$\delta_{\rm{mc}}$ (i.e., the magnetic fluctuation amplitude) 
is unknown and should be determined according to observations. To save the 
computation time, we calculate the mock light curves for a given 
$\delta_{\rm{mc}}$ (denoted as $\delta_{\rm{mc, 0}}$), which is independent 
of $L_{\rm{bol}}$, $M_{\rm{BH}}$, and $\alpha$. The model ensemble squared 
structure functions for other choices of $\delta_{\rm{mc}}$ can be obtained 
by multiplying the factor 
$c=(\delta_{\rm{mc}}/\delta_{\rm{mc, 0}})^2$ with the model squared ensemble 
structure functions for $\delta_{\rm{mc, 0}}$. Finally, we 
calculate the model ensemble squared structure functions ($SF^2_{\rm{mock}}$) 
with the same methodology aforementioned. We repeat the calculation $64$ times 
to estimate the statistical dispersion of a squared structure function.  

%% The "ht!" tells LaTeX to put the figure "here" first, at the "top" next
%% and to override the normal way of calculating a float position
\begin{figure*}
    \includegraphics[width=\columnwidth]{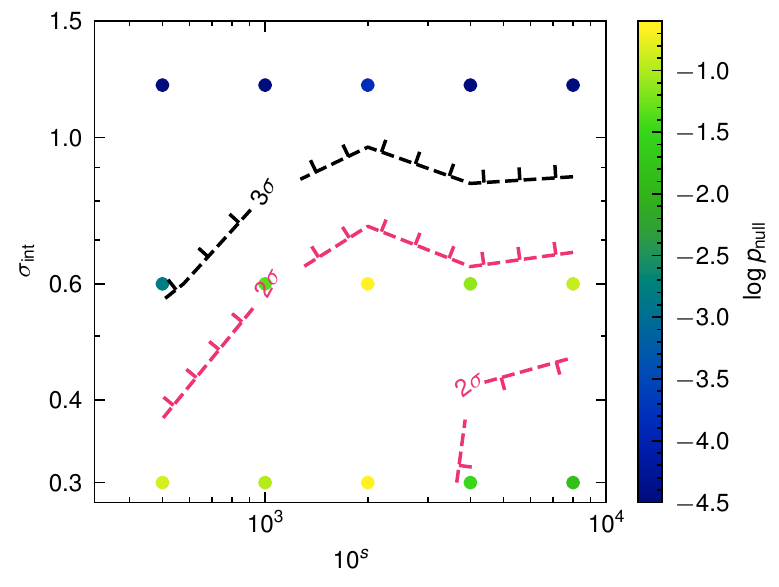}
    \includegraphics[width=\columnwidth]{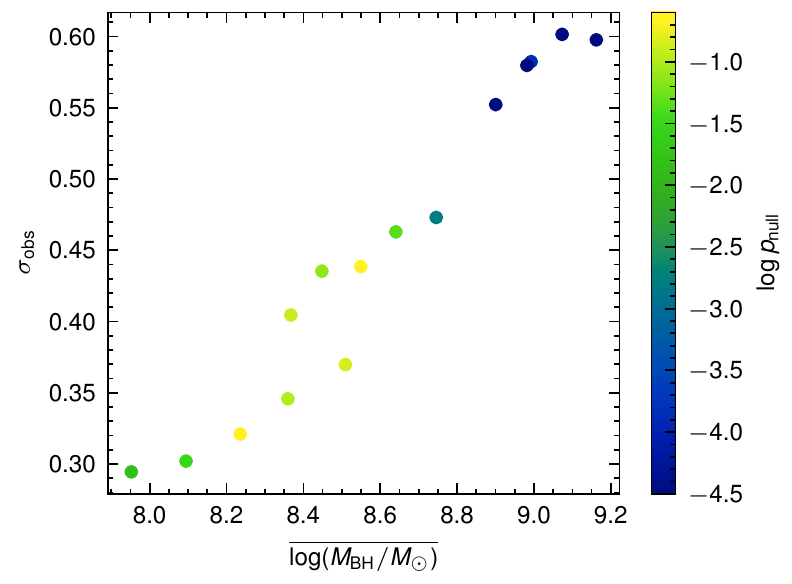}
    \caption{Left: the ``allowed'' normalization and intrinsic scatter of the 
    $M_{\rm{BH}}-M_*$ relation for $\alpha=0.4$. 
    The colours of data points correspond 
    to the null hypothesis probability (i.e., if the model is true, the 
    probability to generate the observed or even larger discrepancy 
    between the data and the model prediction). The black and 
    pink dashed curves indicate the parameter space with 
    $\log p_{\rm{null}}>-1.3$ (i.e., $2 \sigma$) and 
    $\log p_{\rm{null}}>-2.6$ (i.e., $3 \sigma$). No 
    model has $\log p_{\rm{null}}>-0.49$ (i.e., $1\sigma$). 
    Right: The average and NAAD of the $M_{\mathrm{BH}}$ distributions 
    for mock quasars that are ``detectable'' (i.e., brighter 
    than $20.34$ mag in the $i$-band and have broad emission 
    lines). Due to the selection effects, there is a clear correlation 
    between the mean and NAAD of the mass distributions 
    of the ``detectable'' mock quasars. }
    \label{fig:fmass}
\end{figure*}

We use the following statistic for each luminosity bin and each band to fit 
the model against the data and determine $\delta_{\rm{mc}}$, i.e., 
\begin{equation}
    \begin{aligned}
    \chi^2_{\rm{p}, L, \lambda} = {} & \sum \frac{(\log SF_{\rm{obs}}^2 
    - \log(c\times SF_{\rm{mock}}^2 + 
    SF_{\rm{err}}^2))^2}{\sigma_{\rm{tot}}^2} 
    \\ & + \sum \ln (2\pi \sigma_{\rm{tot}}^2) \ ,
    \end{aligned}
\end{equation}
where $SF^2_{\rm{err}}$ represents the squared ensemble 
structure function due to magnitude measurement errors. Note that 
$SF^2_{\rm{err}}$ depends upon $L$ and $\lambda$ (see 
the observed squared structure functions on short timescales in 
Figure~\ref{fig:sfplot}). The variance $\sigma_{\rm{tot}}^2 
= (c^2\sigma_{\rm{SF}}^2+\sigma_{\rm{err}}^2)/(\ln(10)\times 
(c\times SF_{\rm{mock}}^2 + SF_{\rm{err}}^2))^2$, where 
$\sigma^2_{\rm{SF}}$ and $\sigma^2_{\rm{err}}$ represent the 
variance of $SF_{\rm{mock}}^2$ and $SF_{\rm{err}}^2$, 
respectively. We then use \textit{Scipy}'s optimization function to 
find the best-fitting parameters that minimizes 
\begin{equation}
    \chi^2_{\rm{p}}=\sum{\chi^2_{\rm{p}, L, \lambda}} \\.
\end{equation}
There are sixteen free parameters: the first one is $c$, and the remaining 
fifteen ones are for $SF^2_{\rm{err}}$ (five luminosity bins and three 
bands) which are well determined by $SF^2_{\rm{obs}}$ on timescales 
of $\leq 4$ days. 

To measure the goodness of fit, we compute the two-sample Anderson-Darling 
test statistic ($A_{\rm{obs}}$) between the observed and best-fitting 
mock squared structure functions for each luminosity bin and each band. 
We only use data points with $\Delta t>10$ days since the structure 
functions are dominated by measurement errors on shorter timescales. 
Indeed, the squared ensemble structure functions on 
$\Delta t \lesssim 10$ days are larger than those on $\Delta t = 1$ 
day by a factor of $1.2\sim 2.0$ (with a median value of $1.4$), i.e., 
the intrinsic quasar squared ensemble structure functions are 
$20\%\sim 100\%$ of those due to measurement errors. 
Then, we adopt the ``random subset selection'' (RSS) method (which 
is widely used in the interpolated cross-correlation analysis) to assess 
the distribution of the statistic of the Anderson-Darling test. That is, 
we randomly select (with replacement) $N$ data points from an observed 
structure function with $N$ data points; only data points with unique 
$\Delta t$ are used to construct a ``fake'' structure function; if the 
data point of the ``fake'' structure function is larger than ten, we 
apply the two-sample Anderson-Darling test to the observed and ``fake'' 
structure functions and store the corresponding statistic 
($A_{\rm{fake}}$). We repeat these procedures $10, 000$ times. For 
each luminosity bin at each band, we can obtain the NAAD 
of the Anderson-Darling test statistic ($\sigma_A$). The final statistic 
($A_{\rm{obs, tot}}$) between the best-fitting and observed 
structure functions for all luminosity bins and bands is the weighted 
average of $A_{\rm{obs}}$ in each luminosity bin 
and band and the weighting factor is $1/\sigma_A^2$. We can also 
obtain the corresponding fake total statistic 
($A_{\rm{fake, tot}}$), which is the weighted average of 
$A_{\rm{fake}}$, for the $20, 000$ RSS simulations. 
If the best-fitting and observed structure functions are the same (the 
null hypothesis), we expect that the probability ($p_{\rm{null}}$) 
for having $A_{\rm{fake, tot}}\geq A_{\rm{obs, tot}}$ 
in the $10, 000$ RSS simulations is not small. 
Given that we have $20, 000$ RSS simulations, 
if none of $A_{\rm{fake, tot}}$ is larger than $A_{\rm{obs, tot}}$, 
we can conclude that $p_{\rm{null}}\lesssim 0.5/20000=2\times 10^{-5}$. 
We reject models with $p_{\rm{null}}<0.05$ (i.e., $\log p_{\rm{null}} <-1.3$). 

\begin{figure}
    \includegraphics[width=\columnwidth]{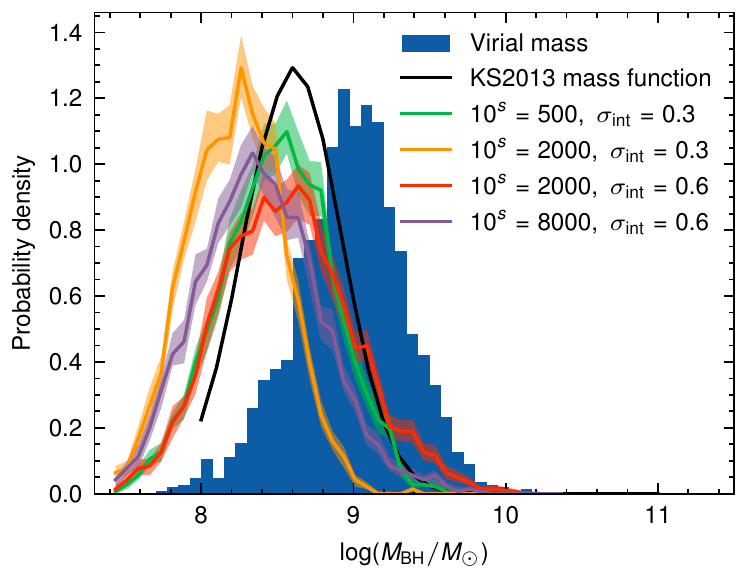}
    \caption{
    The distributions of $\log M_{\rm{BH}}$ for the four cases 
    with the highest $p_{\rm{null}}$ values (the green, yellow, 
    red, and purple curves). The 
    blue-shaded regions represent the distribution of the virial black-hole 
    mass ($\protect \log M_{\rm{BH, vir}}$). The black curve shows the 
    black-hole mass function of KS2013. Our $\protect \log M_{\rm{BH}}$ is, 
    on average, smaller than $\protect \log M_{\rm{BH, vir}}$ by 
    $\protect 0.3\sim 0.6$ dex. The mean $\log M_{\mathrm{BH}}$ 
    values of the models with [$10^s=500$, $\sigma_{\rm{int}}=0.3$ dex] 
    and [$10^s=2000$, $\sigma_{\rm{int}}=0.6$ dex] are slightly smaller 
    than KS2013 by $0.12$ dex and $0.06$ dex, respectively. }
    \label{fig:dism}
\end{figure}

\section{Results}
\label{sect:result}

We compare models with the same $s$ and $\sigma_{\rm{int}}$ 
but different $\alpha$. As an illustration, 
we fix $10^s=500$ and $\sigma_{\rm{int}}=0.3$ 
dex\footnote{The resulting $M_{\mathrm{BH}}$ 
distribution is close to \cite{Kelly2013}, who obtain 
their $M_{\mathrm{BH}}$ distribution by carefully correcting 
for biases in virial black-hole masses.} and present 
the model squared structure functions with $\alpha=0.2$ 
($\alpha=0.4$) in the upper (middle) rows of Figure~\ref{fig:sfplot}. 
The model with $\alpha=0.2$ fits 
the observed squared structure functions poorly and its 
$p_{\rm{null}}$ is much smaller than the alternative model 
with $\alpha=0.4$. In fact, for all the 15 models with $\alpha=0.2$, 
their corresponding $p_{\rm{null}}$ 
are much smaller than $0.04\%$. Hence, the model squared structure 
functions with $\alpha=0.2$ are statistically rejected for these 
luminous quasars (with $L_{\rm{bol}} \gtrsim 10^{45.9}\ 
\rm{erg\ s^{-1}}$). In the subsequent analysis, 
we focus on models with $\alpha=0.4$. 

We can now compare model squared structure functions of various 
$\alpha=0.4$ models with observed ones. The model 
and observed squared structure functions for [$10^s=500$, 
$\sigma_{\rm{int}}=0.3$ dex] and [$10^s=500$, 
$\sigma_{\rm{int}}=1.2$ dex] (both with $\alpha=0.4$) are presented 
in Figure~\ref{fig:sfplot}. Unlike the latter model 
(with $p_{\rm{null}}\lesssim 2\times 10^{-5}$), the former one 
(with $p_{\rm{null}}=0.14$) is not statistically 
rejected by the observations, demonstrating the possibility to 
constrain $M_{\rm{BH}}$ distributions from quasar variability. 

We can calculate the goodness of fit between the observed and model 
squared structure functions for different choices of $s$ and 
$\sigma_{\rm{int}}$ which are the normalization and intrinsic scatter 
of the $M_{\rm{BH}}$-$M_{\star}$ relation, respectively. The results 
(corresponding to $\alpha=0.4$) are presented in 
the left panel of Figure~\ref{fig:fmass}. Cases with 
$\sigma_{\rm{int}}\geq 1.2$ dex have 
$\log p_{\rm{null}}<-2.7$ (i.e., $p_{\rm{null}} < 0.2\%$, which 
is smaller than the probability to be outside the $3\sigma$ regions of 
a Gaussian distribution) and are firmly rejected. Case 1 (i.e., $10^s=500$ 
and $\sigma_{\rm{int}}=0.3$ dex) can explain the observed structure 
functions, which is consistent with recent results of the non-evolution 
of the $M_{\rm{BH}}$--$M_{\star}$ relation \citep[e.g.,][]{Sun2015,Li2021}. 
In the right panel of Figure~\ref{fig:fmass}, we plot the mean 
($\overline{\log (M_{\rm{BH}}/M_{\odot})}$) and NAAD 
($\sigma_{\rm{obs}}$) of the black-hole mass distributions of 
``detectable'' mock quasars for all cases. Due to the selection effects 
(i.e., one can only detect luminous quasars), there is a clear correlation 
between $\overline{\log (M_{\rm{BH}}/M_{\odot})}$ and 
$\sigma_{\rm{obs}}$. The model squared structure functions of cases 
with $\overline{\log (M_{\rm{BH}}/M_{\odot})}= 8.2\sim 8.6$ and 
$\sigma_{\rm{obs}}= 0.3\sim 0.45$ dex are similar to the observed ones 
and have large $p_{\rm{null}}$ (the right panel of 
Figure~\ref{fig:fmass}).  

We show the distributions of $\log M_{\rm{BH}}$ for the four cases 
with highest $p_{\rm{null}}$ (see solid curves with shaded regions 
in Figure~\ref{fig:dism}). 
Comparing with $\log M_{\rm{BH,vir}}$, our $\log M_{\rm{BH}}$ for 
the four cases are systematically smaller by $0.3\sim 0.6$ 
dex.\footnote{In \cite{Sun2020b}, we successfully reproduce the observed 
ensemble structure functions of SDSS S82 light curves with the CHAR 
model and $M_{\rm{BH,vir}}$. This is because the light curves used 
in \cite{Sun2020b} only include SDSS S82 observations and are shorter than 
this work. The SDSS S82 light curves can only probe the ensemble structure 
functions on rest-frame timescales $\lesssim 500$ days for quasars at $z=1.6$. 
On such timescales, models with different $M_{\rm{BH}}$ have similar structure 
functions (see Figure~\ref{fig:sfplot}). Hence, the SDSS 
S82 light curves are too short to constrain $M_{\rm{BH}}$.} This 
systematic offset likely resembles the systematic biases of virial 
black-hole mass as mentioned by \cite[][see their figure 2]{Shen2010}. 
\citet[][hereafter KS2013]{Kelly2013} modelled these biases and inferred 
the bias-corrected 
$\log M_{\rm{BH}}$ distribution from $\log M_{\rm{BH,vir}}$ for SDSS 
detected quasars. The $\log M_{\rm{BH}}$ distributions for the 
two cases with [$10^s=500$, $\sigma_{\rm{int}}=0.3$ dex] and [$10^s=2000$, 
$\sigma_{\rm{int}}=0.36$ dex] are on average smaller than KS2013 
by only $0.12$ dex or $0.06$ dex, respectively. For the other 
two cases, the $\log M_{\rm{BH}}$ distributions are 
on average smaller than that of KS2013 
by no more than $0.4$ dex. Hence, in our opinion, the 
inferred mass distribution is acceptable.

\section{Summary and Future Prospects}
\label{sect:sum}

In this work, we have demonstrated the possibility to infer the $M_{\rm{BH}}$ 
distribution of quasars by modelling their UV/optical multi-wavelength ensemble 
structure functions. Some of our $M_{\rm{BH}}$ distributions 
is roughly consistent with the previous study of \cite{Kelly2013} 
that fully accounts for the statistical biases of $M_{\rm{vir, BH}}$. 
Comparing with existing quasar $M_{\mathrm{BH}}$ measurement methods, 
our approach does not rely on the knowledge of the unknown BLR structure. 
Basing on several assumptions (i.e., all galaxies harbour AGNs, 
$M_{\mathrm{BH}}$ and $M_{\star}$ are correlated, and the distribution of 
$M_{\star}$ and AGN Eddington ratios follow the Schechter function; see 
Section~\ref{sect:mock}), we have also constrained the normalization and 
intrinsic scatter of the possible correlation between $M_{\rm{BH}}$ and 
$M_{\star}$ at redshift $z\simeq 1.6$. 

Compared with SDSS and Pan-STARRS, LSST can provide deeper (by $3\sim 5$ mag) 
and higher cadence time-domain surveys \citep[e.g.,][]{Brandt2018}. Hence, 
its data can extend our analysis to lower black-hole mass ranges. We 
can also explore complex models involving different Eddington ratio 
distributions, the black-hole mass distributions, and SMBH 
spins. The results can also be used to calibrate the virial black hole at 
various redshifts.

\section*{acknowledgments}
We thank J.X. Wang and Y.Q. Xue for helpful discussion. We thank the 
anonymous referee for his/her helpful comments that improved the manuscript. 
M.Y.S. acknowledges support from the National Natural Science Foundation 
of China (NSFC-11973002), the Natural Science Foundation of Fujian Province 
of China (No. 2022J06002), and the China Manned Space Project grant (No. 
CMS-CSST-2021-A06 and CMS-CSST-2021-B11). 

Funding for the SDSS and SDSS-II has been provided by the Alfred P. Sloan 
Foundation, the Participating Institutions, the National Science Foundation, 
the U.S. Department of Energy, the National Aeronautics and Space Administration, 
the Japanese Monbukagakusho, the Max Planck Society, and the Higher Education 
Funding Council for England. The SDSS website is http://www.sdss.org/.

The SDSS is managed by the Astrophysical Research Consortium for the Participating 
Institutions. The Participating Institutions are the American Museum of Natural 
History, Astrophysical Institute Potsdam, University of Basel, University of 
Cambridge, Case Western Reserve University, University of Chicago, Drexel 
University, Fermilab, the Institute for Advanced Study, the Japan Participation 
Group, Johns Hopkins University, the Joint Institute for Nuclear Astrophysics, 
the Kavli Institute for Particle Astrophysics and Cosmology, the Korean 
Scientist Group, the Chinese Academy of Sciences (LAMOST), Los Alamos National 
Laboratory, the Max-Planck Institute for Astronomy (MPIA), the 
Max-Planck-Institute for Astrophysics (MPA), New Mexico State University, Ohio 
State University, University of Pittsburgh, University of Portsmouth, Princeton 
University, the United States Naval Observatory, and the University of Washington. 

The Pan-STARRS1 Surveys (PS1) and the PS1 public science archive have been made 
possible through contributions by the Institute for Astronomy, the University 
of Hawaii, the Pan-STARRS Project Office, the Max-Planck Society and its 
participating institutes, the Max Planck Institute for Astronomy, Heidelberg 
and the Max Planck Institute for Extraterrestrial Physics, Garching, The Johns 
Hopkins University, Durham University, the University of Edinburgh, the Queen's 
University Belfast, the Harvard-Smithsonian Center for Astrophysics, the Las 
Cumbres Observatory Global Telescope Network Incorporated, the National 
Central University of Taiwan, the Space Telescope Science Institute, 
the National Aeronautics and Space Administration under Grant No. NNX08AR22G 
issued through the Planetary Science Division of the NASA Science Mission 
Directorate, the National Science Foundation Grant No. AST-1238877, the 
University of Maryland, Eotvos Lorand University (ELTE), the Los Alamos 
National Laboratory, and the Gordon and Betty Moore Foundation. 

This work makes use of the following Python packages: Matplotlib 
\citep{Hunter2007}, Numpy \& Scipy \citep{scipy} and pyLCSIM 
\citep{Campana2017}.

\section*{Data Availability}
The SDSS S82 light curves can be downloaded from 
\url{https://faculty.washington.edu/ivezic/cmacleod/qso_dr7/Southern.html}. 
The Pan-STARRS second data release site is 
\url{https://catalogs.mast.stsci.edu/panstarrs/}. The CHAR model light 
curves are available upon reasonable request.

\bibliographystyle{mnras}

\renewcommand\thefigure{A\arabic{figure}}

\section*{Appendix A}
\begin{figure*}
    \includegraphics[width=2\columnwidth]{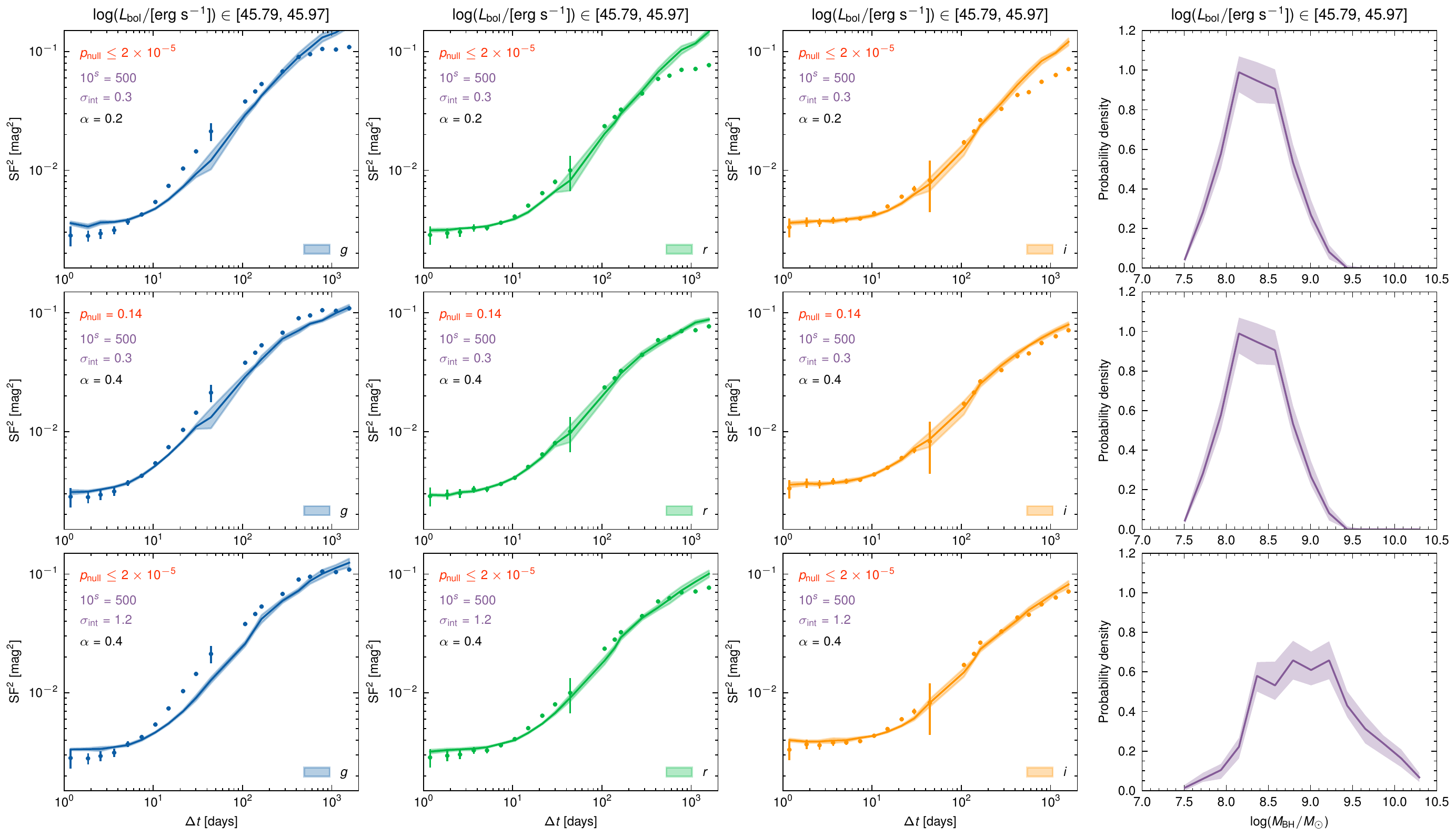}
    \caption{The same as Figure~\ref{fig:sfplot}, but 
    for the luminosity bin $45.79\leq \log (L_{\rm{bol}}/[\rm{erg\ s^{-1}}])<45.97$. 
    \label{fig:a1}}
\end{figure*}

\begin{figure*}
    \includegraphics[width=2\columnwidth]{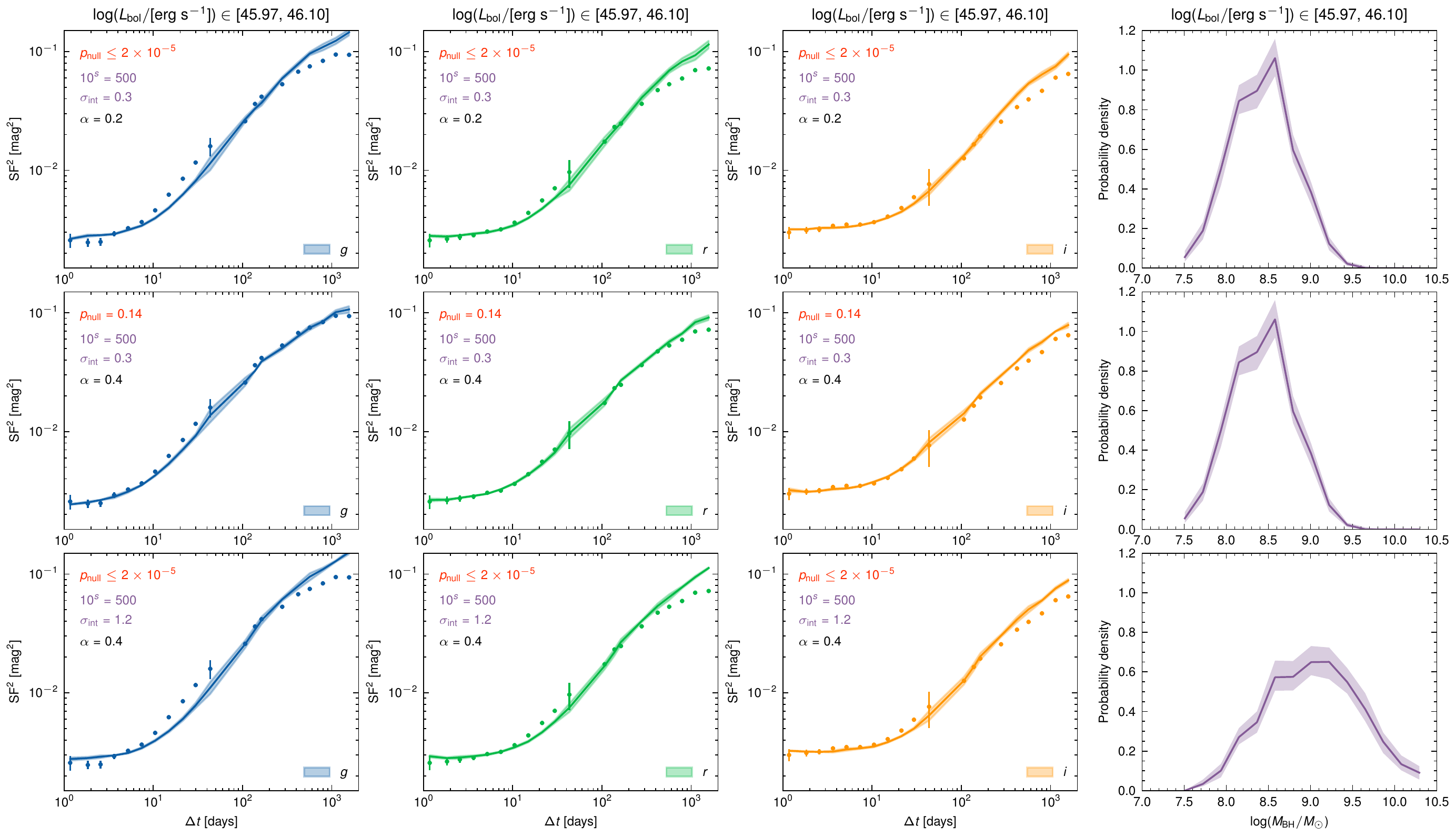}
    \caption{The same as Figure~\ref{fig:sfplot}, but 
    for the luminosity bin $45.97\leq \log (L_{\rm{bol}}/[\rm{erg\ s^{-1}}])<46.10$. 
    \label{fig:a2}}
\end{figure*}

\begin{figure*}
    \includegraphics[width=2\columnwidth]{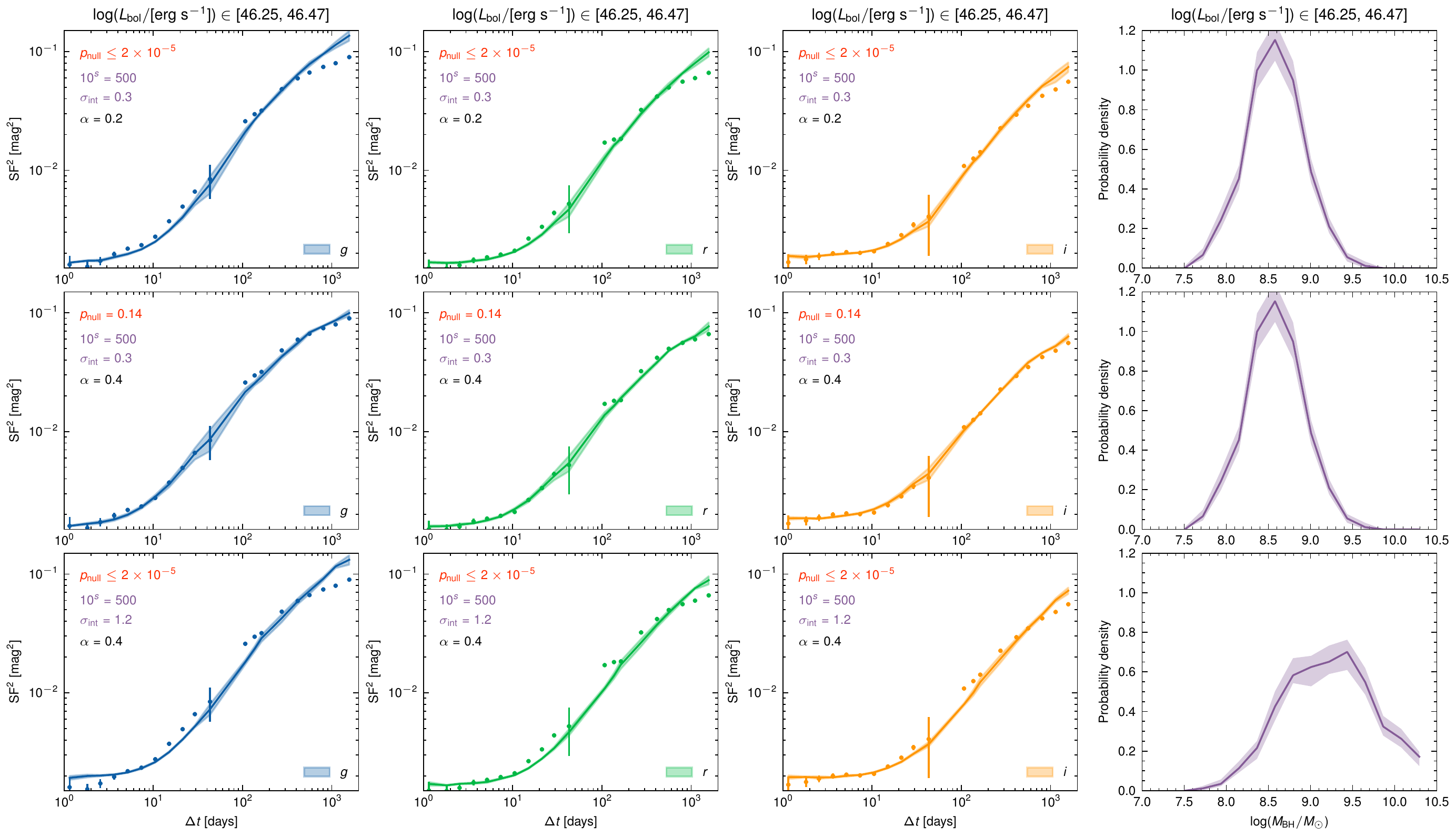}
    \caption{The same as Figure~\ref{fig:sfplot}, but 
    for the luminosity bin $46.25\leq \log (L_{\rm{bol}}/[\rm{erg\ s^{-1}}])<46.47$. 
    \label{fig:a3}}
\end{figure*}

\begin{figure*}
    \includegraphics[width=2\columnwidth]{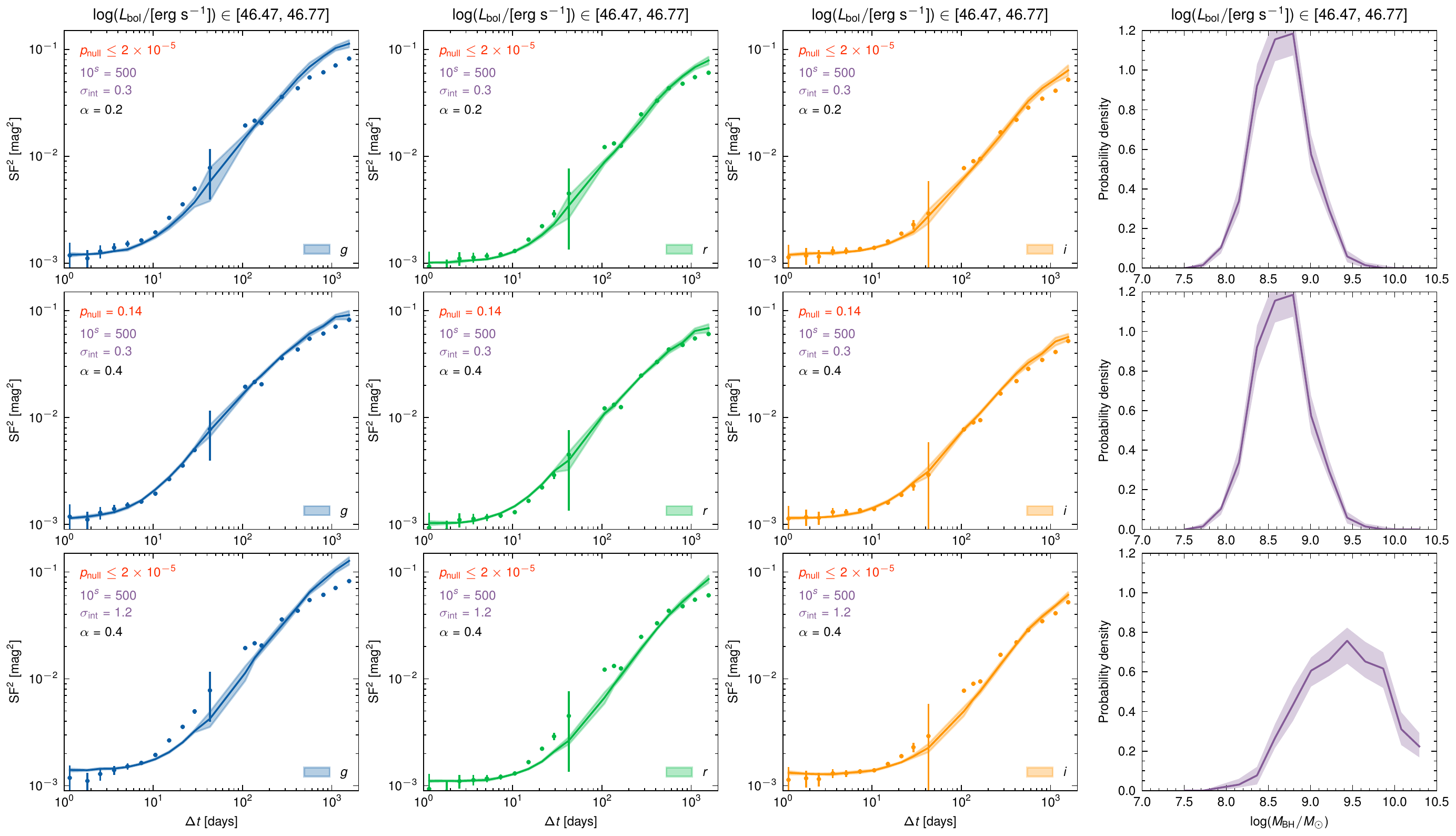}
    \caption{The same as Figure~\ref{fig:sfplot}, but 
    for the luminosity bin $46.47\leq \log (L_{\rm{bol}}/[\rm{erg\ s^{-1}}])<46.77$. 
    \label{fig:a4}}
\end{figure*}

In this section, we illustrate our modelling results for other luminosity 
bins (Figures~\ref{fig:a1},~\ref{fig:a2},~\ref{fig:a3}, and~\ref{fig:a4}). 

% Don't change these lines
\bsp	% typesetting comment
\label{lastpage}
\end{document}